\documentclass[namedreferences]{solarphysics}
\pdfoutput=1
\usepackage[hyperref,optionalrh,solaromanenum]{spr-sola-addons} 
\usepackage{graphicx}                    
\usepackage{color}                       
\usepackage{breakurl}                         
\usepackage{amssymb}
\usepackage{amsbsy}
\usepackage{bm}
\usepackage{latexsym}
\usepackage{marvosym}
\usepackage{enumerate}
\usepackage{textcomp}
\usepackage{float}
\usepackage{times}
\usepackage[varg]{txfonts}

\renewcommand{\vec}[1]{{\boldsymbol #1}}
\newcommand{\etal}{{\it et al.}}

\begin{document}

\begin{article}

\begin{opening}

\title{Can High-Mode Magnetohydrodynamic Waves Propagating in a Spinning Macrospicule Be Unstable due to the Kelvin--Helmholtz Instability?}

%
\author[addressref={aff1},corref,email={izh@phys.uni-sofia.bg}]{\inits{I.~}\fnm{I.~}\lnm{Zhelyazkov}\orcid{orcid.org/0000-0001-6320-7517}}\sep
\author[addressref={aff2}]{\inits{R.~}\fnm{R.~}\lnm{Chandra}\orcid{orcid.org/0000-0002-3518-5856}}

%
\runningauthor{I.~Zhelyazkov and R.~Chandra}
\runningtitle{Kelvin--Helmholtz Instability in a Rotating Macrospicule}

\address[id={aff1}]{Faculty of Physics, Sofia University, 1164 Sofia, Bulgaria}
\address[id={aff2}]{Department of Physics, DSB Campus, Kumaun University, Nainital 263\,001, India}

\begin{abstract}
We investigate the conditions at which high-mode magnetohydrodynamic (MHD) waves propagating in a spinning solar macrospicule can become unstable with respect to the Kelvin--Helmholtz instability (KHI).  We consider the macrospicule as a weakly twisted cylindrical magnetic flux tube moving along and rotating around its axis.  Our study is based on the dispersion relation (in complex variables) of MHD waves obtained from the linearized MHD equations of an incompressible plasma for the macrospicule and cool ($\beta = 0$, rate of the plasma to the magnetic pressure) plasma for its environment.  This dispersion equation is solved numerically at appropriate input parameters to find out an instability region or window that accommodates suitable unstable wavelengths on the order of the macro\-spicule width.  It is established that an $m = 52$ MHD mode propagating in a macro\-spicule with width of $6$~Mm, axial velocity of $75$~km\,s$^{-1}$, and rotating one of $40$~km\,s$^{-1}$ can become unstable against the KHI with instability growth times of $2.2$ and $0.57$~min at $3$ and $5$~Mm unstable wavelengths, respectively.  These growth times are much shorter than the macrospicule lifetime, which lasts about $15$~min.  An increase or decease in the width of the jet would change the KHI growth times, which remain more or less on the same order when they are evaluated at wavelengths equal to the width or radius of the macrospicule.  It is worth noting that the excited MHD modes are super-Alfv\'enic waves.  A change in the background magnetic field can lead to another MHD mode number $m$ that ensures the required instability window.
\end{abstract}

%

\end{opening}

%
\section{Introduction}
\label{sec:intro}
Macrospicules were first detected four decades ago by \citet{bohlin1975} from He~\textsc{ii} $304$~\AA{} spectroheliograms, that were obtained with the \emph{Naval Research Laboratory Ext\-reme-Ultra\-violet Spectrograph\/} during a \emph{Skylab\/} mission.  Macrospicules are usually seen as jets $5$--$15^{\prime \prime}$ in diameter, and $5$--$50^{\prime \prime}$ in length that move outward into the corona at speeds of $10$--$150$~km\,s$^{-1}$ and have lifetimes of $5$--$45$~min.  An essential step in observing macrospicules was made using the \emph{Coronal Diagnostic Spectrometer\/} (CDS) on board the \emph{Solar and Heliospheric Observatory\/} (SOHO: \citealp{domingo1995}).  First scientific results on using CDS were reported during 1997, and an extensive review of the obtained physical parameters of the observed macrospicules can be found in \citet{harrison1997}.  \citet{pike1998} analyzed macrospicule observations from 9 April 1996 to 4 April 1997 and first reported that macro\-spicules can be considered as spinning columns with rotating velocity of $20$--$50$~km\,s$^{-1}$.  \citet{banerjee2000} used SOHO \emph{Extreme ultraviolet Imaging Telescope\/} (EIT) images in the O~\textsc{v} $629$~\AA{} line that was detected a giant macrospicule at the limb on 15 July 1999.  The authors were able to follow its dynamical structure.  \citet{parenti2002} detected from the analysis of a sequence of SOHO/CDS observations that were obtained off-limb in the south polar coronal hole on 6 March 1998, a jet-like feature that was visible in the chromospheric and low transition region lines, which turned out to be a macrospicule.  According to these authors, the macrospicule was found to have a density of the order on $10^{10}$~cm$^{-3}$ and a temperature of about $2$--$3 \times 10^5$~K.  The initial outflow velocity near the limb was over $80$~km\,s$^{-1}$.  \citet{kamio2010} used \emph{Hinode\/} \citep{kosugi2007} \emph{Extreme-ultraviolet Imaging Spectrometer\/} (EIS: \citealp{culhane2007}) and the \emph{Solar Ultraviolet Measurements of Emitted Radiation instrument\/} (SUMER:  \citealp{wilhelm1995}) on board SOHO to measure the line-of-sight (LOS) motions of both macrospicule and coronal jets.  At the same time, with the help of the \emph{X-Ray Telescope\/} (XRT: \citealp{golub2007}) on board \emph{Hinode\/} and \emph{Sun--Earth Connection Coronal and Heliospheric Investigation\/} (SECCHI) instrument suite \citep{howard2008} on the \emph{Solar Terrestrial Relations Observatory\/} (STEREO: \citealp{kaiser2008}), the authors traced the evolution of the coronal jet and the macrospicule.  The upward-propagating and rotating velocities of the macrospicule, averaged over $10$~min, between 02:36 UT and 02:46 UT, proved to be $130 \pm 30$ and $25 \pm 5$~km\,s$^{-1}$, respectively.  \citet{scullion2010} explored the nature of macrospicule structures, both off-limb and on-disk, using the high-resolution spectroscopy obtained with the SOHO/SUMER instrument.  The authors reported on finding high-velocity features observed simultaneously in spectral lines formed in the mid-transition region, N~\textsc{iv} $765$~\AA{} spectral line ($1.4 \times 10^5$~K), and in the low corona, Ne~\textsc{viii} $770$~\AA{} spectral line ($6.3 \times 10^5$~K), while in the hot Ne~\textsc{viii} $770$~\AA{} line the flow speed reached ${\approx}145$~km\,s$^{-1}$.  \citet{madjarska2011} performed multi-instrument observations with SUMER/SOHO and with the EIS/SOT/XRT/\emph{Hinode\/} at the north pole on 28 and 29 April 2009 and detected three rotating macrospicules.  Their main result is that although very large and dynamic, these spicules do not appear in spectral lines that formed at temperatures above $300\,000$~K.  In the same year, 2011, \citet{murawski2011} presented the first numerical simulation of macrospicule formation by implementing the VAL-C model of solar temperature \citep{vernazza1981}.  Using the FLASH code, these authors solved the two-dimensional ideal MHD equations to model a macrospicule, whose physical parameters match those of a solar spicule observed at the north polar region in the $304$~\AA{} of the \emph{Atmospheric Imaging Assembly\/} (AIA: \citealp{lemen2012}) on board the \emph{Solar Dynamics Observatory\/} (SDO: \citealp{pesnell2012}) on 3 August 2010.  The essence of this numerical simulation is that the solar macrospicules can be triggered by velocity pulses launched from the chromosphere.  Another mechanism for the origin of macrospicules was proposed by \citet{kayshap2013}, who numerically modeled the triggering of a macrospicule and a jet observed by AIA/SDO on 11 November 2010 in the north polar corona.  The spicule, which was considered to be a magnetic flux tube that undergoes kinking, reached up to ${\approx}40$~Mm in the solar atmosphere with a projected speed of ${\approx}95$~km\,s$^{-1}$.  The simulation results, obtained in the same manner as in \citealp{murawski2011}, show that a reconnection-generated velocity pulse in the lower solar atmosphere steepens into a slow shock and the cool plasma is driven behind it in the form of a macrospicule.  \citet{loboda2017} used high-cadence EUV observations obtained by the Lebedev Institute of Physics TESIS solar observatory, which is a set of five space instruments, to perform a detailed investigation of the axial plasma motions in solar macrospicules.  They used a one-dimensional hydrodynamic method to reconstruct the evolution of the internal velocity field of 18 macrospicules and found that 15 of them followed parabolic trajectories with high precision, which corresponds closely to the obtained velocity fields.  Two articles, namely those of \citet{bennett2015} and \citet{kiss2017}, summarized the origin, evolution, and physical parameters of a large number of observed jets: $101$ macrospicules in \citet{bennett2015}, and $301$ in \citet{kiss2017}.  The first dataset covers observations over a $2.5$-year-long time interval, while the second database is built on the observations of macrospicules that occurred between June 2010 and December 2015, that is, over a $5.5$-year-long time interval.  Table~2 in \citet{kiss2017} contains the main macrospicule characteristics obtained from `old' and `new' observations.  Some averaged values are a lifetime of $16.75 \pm 4.5$~min, a maximum width of $6.1 \pm 4$~Mm, and an average upflow velocity of $73.14 \pm 25.92$~km\,s$^{-1}$.  According to the authors, the maximum length is overestimated, having a value of $28.05 \pm 7.67$~Mm.

Solar spicules, like every magnetically structured entity in the solar atmosphere, support the propagation of different types of MHD waves---for a review of a large number of observational and theoretical investigations of oscillations and waves in spicules see, for instance, \cite{zaqarashvili2009}.  The axial mass flow in spicules can be the reason that causes the propagating MHD modes to become unstable due to a velocity jump at the spicule surface and the instability that occurs is of the Kelvin--Helmholtz kind.  We recall that the Kelvin--Helmholtz instability (KHI) is a purely hydrodynamic phenomenon and it arises at the interface of two fluid layers that move with different speeds (see, \emph{e.g.}, \citealp{chandrasekhar1961}); then a strong velocity shear comes into being near the thin interface region of these two fluids forming a vortex sheet that becomes unstable to the spiral-like perturbations at small spatial scales.  It is worth noting that the magnetic field plays an important role in the instability that develops, notably, a strong enough magnetic filed can suppress the KHI.  Solar spicules are usually modeled as moving untwisted cylindrical magnetic flux tubes and the vortex sheet emerging near the tube boundary may become unstable against the KHI provided that its axial velocity exceeds some critical or threshold value \citep{ryu2000}.  This vortex sheet, during the nonlinear stage of the KHI, causes the conversion of the directed flow energy into a turbulent energy, which creates an energy cascade at smaller spatial scales.  First theoretical modelings of the KHI in the relatively faster Type-II spicules \citep{depontieu2007} (see, \emph{e.g.}, \citealp{zhelyazkov2012}; \citealp{zhelyazkov2013}; \citealp{ajabshirizadeh2015}; \citealp{ebadi2016}) show that the magnitudes of the threshold flow velocities at which the KHI rises critically depend on the density contrast defined as $\rho_\mathrm{e}/\rho_\mathrm{i} = \eta$, where $\rho_\mathrm{i}$ is the spicule plasma density, and $\rho_\mathrm{e}$ is that of the surrounding magnetized plasma.  The study of the KHI in a spicule with $\eta = 0.01$ (\citealp{zhelyazkov2012}; \citealp{zhelyazkov2013}) shows that the critical velocity for the instability onset of the kink ($m = 1$) MHD mode must be higher than $876$~km\,s$^{-1}$, which is generally not accessible for Type-II spicules.  A decrease in the density contrast, say, taking $\eta = 0.02$, leads to a decrease of the threshold speed to $711$~km\,s$^{-1}$, which is still too high.  \citet{ajabshirizadeh2015} explored the KHI of the kink ($m = 1$) mode in Type-II spicules at much lower value of the density contrast (${=}0.1$) and also claimed that the required threshold flow velocity is higher than the velocity assumed by them, which is $150$~km\,s$^{-1}$ (see also a comment to their results in \citealp{zhelyazkov2016}).  A two-dimensional MHD numerical modeling (using the Athena3D code) of the KHI in solar spicules performed by \citet{ebadi2016} shows that at some selected densities and flow speeds, a KHI-type onset and transition to turbulent flow in spicules can be observed.

Recent studies (\citealp{depontieu2014}; \citealp{iijima2017}) showed that all small-scale jets in the solar chromosphere and transition region should possess some magnetic field twist.  In particular, \citet{iijima2017} performed a three-dimensional MHD simulation of the formation of solar chromospheric jets with twisted magnetic field lines and were able to produce a tall chromospheric jet with a maximum height of $10$--$11$~Mm and a lifetime of $8$--$10$~min.  These authors also found that the produced chromospheric jet forms a cluster with a diameter of several Mm with finer strands, and they claimed that the obtained results imply a close relationship between the simulated jet and solar spicules.  The magnetic field twist might weakly change the value of the critical flow velocity required for instability onset of the kink ($m = 1$) MHD mode propagating along Type-I or Type-II spicules, but it will hardly dramatically diminish it; this has to be checked.  The case of spinning magnetically twisted macrospicules is more specific---as \citet{zaqarashvili2015} have established, in axially moving solar jets that rotate around their axes, the KHI can only be excited at high ($m \geqslant 2$) MHD modes.  The aim of our study is to see at what wave mode number $m$ the occurrence of a KHI can be expected in a rotating macrospicule at wavelengths of the unstable mode comparable to its width or radius.  As a model we use the macrospicule observed on 8 March 1997 at 00:02 UT \citep{pike1998}.  The axial velocity of that jet is $75$~km\,s$^{-1}$, and we evaluate its rotating speed to be $40$~km\,s$^{-1}$.  In addition, we investigate how the magnitude of the jet width influences the KHI characteristics of the excited high MHD mode.

Our article is organized as follows.  In Section~2, we present the observational data.  The magnetic field topology of a moving and rotating cylindrical flux tube modeling the solar macrospicule, along with its physical parameters and the normal mode dispersion relation of the excited wave are given in Section~3.  Section~4 is devoted to the numerical solutions to the wave dispersion relation and to the discussion of the obtained results.  The new findings and comments on future improvements of KHI studies in solar spinning macrospicules can be found in the last section.

\section{Observations}
\label{sec:observations}
\citet{pike1998} reported a statistical study of the dynamics of solar transition region features, such as macrospicules.  These features were observed on the solar disk as well as on the solar limb.  For their investigation, as we described in Section~1, they used the data from the \emph{Coronal Diagnostic Spectrometer\/} (CDS) on board SOHO.  In their article, the authors discussed the unique CDS observations of a macrospicule that was first reported by \citet{pike1997} along with their own (Pike and Mason) observations from the \emph{Normal Incidence Spectrometer\/} (NIS), which covers the wavelength range from $307$ to $379$~\AA{} and that from $513$ to $633$~\AA{} using a microchannel plate and CCD combination detector.  The details of macrospicule events observed near the limb are given in Table I in \citealp{pike1998}, while those of macrospicule events observed on the disk are presented in Table II.  The main finding of their study was the rotation in these features.  Their conclusion was based on the red- and blueshifted emission on either side of the macrospicule axes and the detected rotation assuredly plays an important role in the dynamics of the transition region.  Our choice for modeling the event observed on 8 March 1997 at 00:02~UT (see Table II in \citealp{pike1998}) is the circumstance that that macrospicule possesses, more or less, the basic characteristics of the tornado-like jets that have been observed over the years.

\section{Geometry, Physical Parameters, and Wave Dispersion Relation}
\label{sec:geometry}
Our model of the macrospicule is similar to that in our previous articles \citet{zhelyazkov2018a} and \citet{zhelyazkov2018b}; a macrospicule is considered as a cylindrical weakly twisted magnetic flux tube with radius $a$ and homogeneous density $\rho_\mathrm{i}$ moving with velocity $\vec{U}$.  The tube is surrounded by a plasma with homogeneous density $\rho_\mathrm{e}$ being embedded in an homogeneous magnetic field $\vec{B}_\mathrm{e}$, which in cylindrical coordinates ($r, \phi, z$), has only a $z$ component, that is,  $\vec{B}_\mathrm{e} = (0, 0, B_\mathrm{e})$.  (In our article the label `i' is the abbreviation for interior, while the label `e' means exterior.)  The internal magnetic field and the flow velocity, by contrast, are both twisted and can be generally represented by the vectors $\vec{B}_\mathrm{i} = \left( 0, B_{\mathrm{i}\phi}(r), B_{\mathrm{i}z} \right)$ and $\vec{U} = \left( 0, U_{\phi}(r), U_{z} \right)$, respectively.  We note that the $B_{\mathrm{i}z}$ and $U_z$ are constant.  Concerning the azimuthal magnetic and flow velocity components, we assume that they are linear functions of the radial position $r$, and evaluated at the tube interface they are constants, equal to $B_{\phi} = Aa$ and $U_{\phi} = \Omega a$, respectively.  In particular, $\Omega$ is the macrospicule angular speed.  Thus, in equilibrium the rigidly rotating, constant-magnetic-pitch column that models the macrospicule, should satisfy the force-balance equation (see, \emph{e.g.}, \citealp{chandrasekhar1961}; \citealp{goossens1992})
\begin{equation}
\label{eq:forceeq}
    \frac{\mathrm{d}}{\mathrm{d}r}\left( p_\mathrm{i} + \frac{B_\mathrm{i}^2}{2\mu} \right) = \frac{\rho_\mathrm{i}U_\phi^2}{r} - \frac{B_{\mathrm{i}\phi}^2}{\mu r},
\end{equation}
where $\mu$ is the plasma permeability and $p_\mathrm{t} = p_\mathrm{i} + B_\mathrm{i}^2/2\mu$ is the total (thermal plus magnetic) pressure, in which $B_\mathrm{i}^2 = B_{\mathrm{i}\phi}^2(r) + B_{\mathrm{i}z}^2$.  The above equation says that the total pressure radial gradient should balance the centrifugal force and the force due to the curved magnetic field lines.  Upon integrating Equation~\ref{eq:forceeq} from $0$ to the tube radius $a$, bearing in mind the linear dependence of both $U_\phi$ and $B_{\mathrm{i}\phi}$ on $r$, we obtain that
\[
    p_\mathrm{t}(a) = p_\mathrm{t}(0) + \frac{1}{2}\rho_\mathrm{i}U_\phi^2(a) - \frac{B_{\mathrm{i}\phi}^2(a)}{2\mu},
\]
where $p_\mathrm{t}(0) = p_\mathrm{i}(0) + B_{\mathrm{i}z}^2/2\mu$.  (The integration can in principle  be performed from $0$ to any $r$ to obtain the radial profile of the total pressure inside the tube---for an equivalent expression of $p_\mathrm{t}(r)$, derived from integration of the momentum equilibrium equation for the equilibrium variables, see Equation~(2) in \citealp{zhelyazkov2018a}.)  The above internal total pressure, $p_\mathrm{t}(a)$ (evaluated at the tube boundary), must be balanced by the total pressure of the surrounding plasma which implies that
\[
    p_\mathrm{t}(0) + \frac{B_{\mathrm{i}z}^2}{2\mu} - \frac{B_\phi^2}{2\mu} + \frac{1}{2}\rho_\mathrm{i}U_\phi^2 = p_\mathrm{e} + \frac{B_{\mathrm{e}}^2}{2\mu}.
\]
We recall that $B_\phi = B_{\mathrm{i}\phi}(a)$.  The obtained total pressure balance equation can be presented in the form
\begin{equation}
\label{eq:pbeq}
    p_\mathrm{i} + \frac{1}{2}\rho_\mathrm{i}U_\phi^2 + \frac{B_{\mathrm{i}z}^2}{2\mu}\left( 1 - \varepsilon_1^2 \right) = p_\mathrm{e} + \frac{B_\mathrm{e}^2}{2\mu},
\end{equation}
where $p_\mathrm{i}$ is the thermal pressure at the tube axis and $\varepsilon_1 \equiv B_{\phi}/B_{\mathrm{i}z} = Aa/B_{\mathrm{i}z}$ is the magnetic field twist parameter.  In a similar way we introduce $\varepsilon_2 \equiv U_\phi/U_z$, which characterizes the jet velocity twist.  In the above equation, $p_\mathrm{e}$ denotes the thermal pressure in the environment.  The choice of plasma and environment parameters must be such that the total pressure balance Equation~\ref{eq:pbeq} is satisfied.  It is important to note that in our case $\varepsilon_2$ is defined by observationally measured rotational and axial velocities while $\varepsilon_1$ is a parameter that has to be specified when using Equation~\ref{eq:pbeq}.

Since \citet{pike1998} did not provide any data concerning the macro\-spicule and its environment electron number densities, $n_\mathrm{i}$ and $n_\mathrm{e}$, respectively, based on measurements of  chromospheric jets, we assume that $n_\mathrm{i} = 1.0 \times 10^{10}$~cm$^{-3}$ and $n_\mathrm{e} = 1.0 \times 10^{9}$~cm$^{-3}$ to have a jet that is at least one order denser than the surrounding plasma.  We take the macrospicule temperature to be $T_\mathrm{i} = 5.0 \times 10^5$~K, while that of its environment is typically equal to $1$~MK, that is, $T_\mathrm{e} = 1.0 \times 10^6$~K.  This choice of electron number densities and electron temperatures defines the density contrast $\eta \equiv n_\mathrm{e}/n_\mathrm{i} = 0.1$, and the sound speeds in both media: $c_\mathrm{si} = 83.0$ and $c_\mathrm{se} = 117.3$~km\,s$^{-1}$, respectively.  Assuming a relatively weak internal magnetic field twist $\varepsilon_1 = 0.005$ and a background magnetic field $B_\mathrm{e} = 5$~G, from the total pressure balance Equation \ref{eq:pbeq}, we obtain the Alfv\'en speeds $v_\mathrm{Ai} = 60.6$ and $v_\mathrm{Ae} = 344.7$~km\,s$^{-1}$, respectively, the ratio of axial magnetic fields, $b \equiv B_\mathrm{e}/B_{\mathrm{i}z} = 1.798$, as well as the two plasma betas, $\beta_\mathrm{i} = 2.248$ and $\beta_\mathrm{e} = 0.139$.  We note that the Alfv\'en speed inside the macrospicule is defined (and computed) as $v_\mathrm{Ai} = B_{\mathrm{i}z}/\sqrt{\mu \rho_\mathrm{i}}$, while both plasma betas are evaluated from the ratio $c_\mathrm{s}^2/v_\mathrm{A}^2$ (multiplied by $6/5$), where the Alfv\'en speeds are calculated with the full magnetic fields.  The basic physical parameters of the macrospicule and its environment are summarized in Table~\ref{tab:parameters}.
\begin{table}
\caption{Macrospicule and its environment physical parameters derived at a background magnetic field $B_\mathrm{e} = 5$~G.}
\label{tab:parameters}
\vspace*{2mm}


\begin{tabular}{cccc}
\hline
Medium & Temperature & Electron density & Plasma beta \\
       & (MK)        &  (${\times}10^{10}$ cm$^{-3}$) & \\
\hline
Macrospicule & $0.5$ & $1.0$ & $2.248$  \\
Environment  & $1.0$ & $0.1$ & $0.139$  \\
\hline
\end{tabular}
\end{table}
We recall that the macrospicule axial speed is $U_z = 75$~km\,s$^{-1}$, while its rotational one is $U_\phi = 40$~km\,s$^{-1}$.  We assume that the macrospicule width is $\Delta \ell = 6$~Mm, its height $H = 28$~Mm, and the lifetime is on the order of $15$~min.  All these data are more or less similar to the averaged macrospicule parameters discussed in \citealp{kiss2017}.

The dispersion relation of high-mode ($m \geqslant 2$) MHD waves traveling in a magnetized axially moving and rotating jet in \citet{zaqarashvili2015} was obtained under the assumption that both media (the jet and its environment) are incompressible plasmas.  As seen from Table~\ref{tab:parameters}, the macrospicule plasma beta is larger than $1$ and the jet medium can be treated as a nearly incompressible fluid \citep{zank1993}.  On the other hand, the plasma beta of the surrounding magnetized plasma is lower than $1$ and it is more adequate to consider it as a cool medium.  This implies that the wave dispersion relation, derived in \citet{zaqarashvili2015}, has to be slightly modified.  We skip the derivation of this modified equation from the basic MHD equations, which has been done in \citet{zhelyazkov2018a}.  Here we provide its final form---the full derivation of the wave dispersion relation is presented in the Appendix.  As is logical to expect, the equation has to be expressed in terms of modified Bessel functions of first and second kind, $I_m$ and $K_m$, and their derivatives, $I^\prime_m$ and $K^\prime_m$, with respect to the functions arguments $\kappa_\mathrm{i}a$ and $\kappa_\mathrm{e}a$, respectively, that is,
\begin{eqnarray}
\label{eq:dispeq}
    \frac{\left( \sigma^2 - \omega_\mathrm{Ai}^2 \right)F_m(\kappa_\mathrm{i}a) - 2m\left( \sigma \Omega + A\omega_\mathrm{Ai}/\! \sqrt{\mu \rho_\mathrm{i}} \right)}{\rho_\mathrm{i}\left( \sigma^2 - \omega_\mathrm{Ai}^2 \right)^2 - 4\rho_\mathrm{i}\left( \sigma \Omega + A\omega_\mathrm{Ai}/\! \sqrt{\mu \rho_\mathrm{i}} \right)^2} \nonumber \\
    \nonumber \\
    {}= \frac{P_m(\kappa_\mathrm{e} a)}{\rho_\mathrm{e}\left( \sigma^2 - \omega_\mathrm{Ae}^2 \right) - \left( \rho_\mathrm{i}\Omega^2 - A^2/\mu \right)P_m(\kappa_\mathrm{e} a)},
\end{eqnarray}
where $\Omega$ is the macrospicule angular velocity, $A$ is a constant, defining the linear radial profile of the azimuthal magnetic field component, $B_{\mathrm{i}\phi}$, and
\[
    F_m(\kappa_\mathrm{i}a) = \frac{\kappa_\mathrm{i}aI_m^{\prime}(\kappa_\mathrm{i}a)}{I_m(\kappa_\mathrm{i}a)} \quad \mathrm{and} \quad P_m(\kappa_\mathrm{e} a) = \frac{\kappa_\mathrm{e} aK_m^{\prime}(\kappa_\mathrm{e} a)}{K_m(\kappa_\mathrm{e} a)}.
\]
Here,
\[
    \kappa_\mathrm{i}^2 = k_z^2\left[ 1 - 4\left( \frac{\sigma \Omega + A\omega_\mathrm{Ai}/\!\sqrt{\mu \rho_\mathrm{i}}}{\sigma^2 - \omega_\mathrm{Ai}^2} \right)^2 \right] \quad \mathrm{and} \quad \kappa_\mathrm{e}^2 = k_z^2 \left( 1 - \frac{\omega^2}{\omega_\mathrm{Ae}^2}\right)
\]
are the squared wave amplitude attenuation coefficients in both media, in which
\[
    \omega_\mathrm{Ai} = \left( \frac{m}{r}B_{\mathrm{i}\phi} + k_z B_{\mathrm{i}z} \right)/\sqrt{\mu \rho_\mathrm{i}} \quad \mathrm{and} \quad \omega_\mathrm{Ae} = k_z B_\mathrm{e}/\sqrt{\mu \rho_\mathrm{e}}
\]
are the corresponding local Alfv\'en frequencies, and
\[
    \sigma = \omega - \frac{m}{r}U_{\phi} - k_z U_z
\]
is the Doppler-shifted wave frequency in the macrospicule.  The solutions to the dispersion relation (Equation~\ref{eq:dispeq}) are given and used in the next section.

\section{Numerical Results and Discussion}
\label{sec:numerics}
Because we search for an instability of MHD high ($m \geqslant 2$) modes in the macrospicule--coronal plasma system, we assume that the angular wave frequency, $\omega$, is a complex quantity, while the propagating wave number, $k_z$, is a real quantity.  To perform the numerical task, we normalize the velocities with respect to the Alfv\'en speed inside the macrospicule, $v_\mathrm{Ai}$, and the lengths with respect to $a$ (the tube radius).  The normalization of Alfv\'en local and Doppler-shifted frequencies along with the Alfv\'en speed in the environment requires the usage of both twist parameters, $\varepsilon_1$, $\varepsilon_2$, as well as the magnetic fields ratio $b = B_\mathrm{e}/B_{\mathrm{i}z}$.  The normalized axial flow velocity is presented by the Alfv\'en Mach number $M_\mathrm{A} = U_z/v_\mathrm{Ai}$.  Thus, the input parameters in the numerical solving of the transcendental Equation~\ref{eq:dispeq} (in complex variables) are: $m$, $\eta$, $\varepsilon_1$, $\varepsilon_2$, $b$, and $M_\mathrm{A}$.  It has been derived in \citealp{zaqarashvili2015} that the instability in an untwisted rotating flux tube at sub-Alfv\'enic jet velocities can occur if
\begin{equation}
\label{eq:criterion}
    \frac{a^2 \Omega^2}{v_\mathrm{Ai}^2} > \frac{1 + \eta}{1 + |m|\eta}\,\frac{(k_z a)^2}{|m| - 1}(1 + b^2).
\end{equation}
As seen, each rotating jet can be unstable for any mode number $m \geqslant 2$.  This inequality is applicable to slightly twisted spinning jets provided that the magnetic field twist $\varepsilon_1$ is relatively small, say between $0.001$ and $0.005$.  Here, we make an important assumption that the axial velocity of the macrospicule, $U_z$, deduced from observations is the threshold speed for the KHI onset.  Then, for fixed values of $m$, $\eta$, $U_\phi = \Omega a$, $v_\mathrm{Ai}$, and $b = B_\mathrm{e}/B_{\mathrm{i}z}$, the above inequality defines the right-hand-side limit of the instability range on the $k_z a$-axis
\begin{equation}
\label{eq:instcond}
    (k_z a)_\mathrm{rhs} < \left\{ \left( \frac{U_\phi}{v_\mathrm{Ai}} \right)^2 \frac{1 + |m|\eta}{1 + \eta}\,\frac{|m| - 1}{1 + b^2} \right\}^{1/2}.
\end{equation}
This inequality says that the instability can occur for all dimensionless wavenumbers $k_z a$ lower than $(k_z a)_\mathrm{rhs}$.  However, one can talk of instability if the unstable wavelength is shorter than the height of the jet.  This requirement allows us to define the left-hand-side limit of the instability region:
\begin{equation}
\label{eq:lhlimit}
    (k_z a)_\mathrm{lhs} > \frac{\pi \Delta \ell}{H}.
\end{equation}
In our case this limit is equal to $0.673$.  Numerical computations show that for small MHD mode numbers, $m$, the instability range or window is relatively narrow and the shortest unstable wavelengths, $\lambda_\mathrm{KH} = \pi\,\Delta \ell/k_z a$, that can be computed are much larger than the macrospicule width.  Such long wavelengths are not comfortable for instability detection or observation.  The rapidly developed vortex-like structures at the boundary of the jet have the size of the width or radius of the flux tube (see, \emph{e.g.}, Figure~1 in \citealp{zhelyazkov2018a}).  Thus, we should look for such instability regions that would contain the expected unstable wavelengths.  A noticeable extension of the instability range can be achieved via increasing the wave mode number.  If we wish, for example, to have an unstable wavelength $\lambda_\mathrm{KH} = 3$~Mm (the half-width of our macrospicule), we have to find out that mode number $m$ whose instability window will accommodate $k_z a = 2 \pi$ (the dimensionless wavenumber that corresponds to $\lambda_\mathrm{KH} = 3$~Mm).  An estimation of the required mode number for an $\varepsilon_1 = 0.005$ rotating flux tube can be obtained by presenting the instability criterion in Equation~\ref{eq:criterion} in the form
\begin{equation}
\label{eq:findingm}
    \eta |m|^2 + (1 - \eta)|m| - 1 - \frac{(k_z a)^2(1 + \eta)(1 + b^2)}{(U_\phi/v_\mathrm{Ai})^2} > 0.
\end{equation}
With $\eta = 0.1$, $k_z a = 2\pi$, $U_\phi = 40$~km\,s$^{-1}$, $v_\mathrm{Ai} = 60.6$~km\,s$^{-1}$, and $b = 1.798$, the above equation yields $m = 64$.  This magnitude is, however, overestimated---our numerical calculations show that the appropriate MHD wave mode number that accommodates the unstable wavelength of $3$~Mm ($k_z a = 2\pi$) is $m = 52$.

\begin{figure}[!ht]
   \centerline{\hspace*{0.015\textwidth}
               \includegraphics[width=0.515\textwidth,clip=]{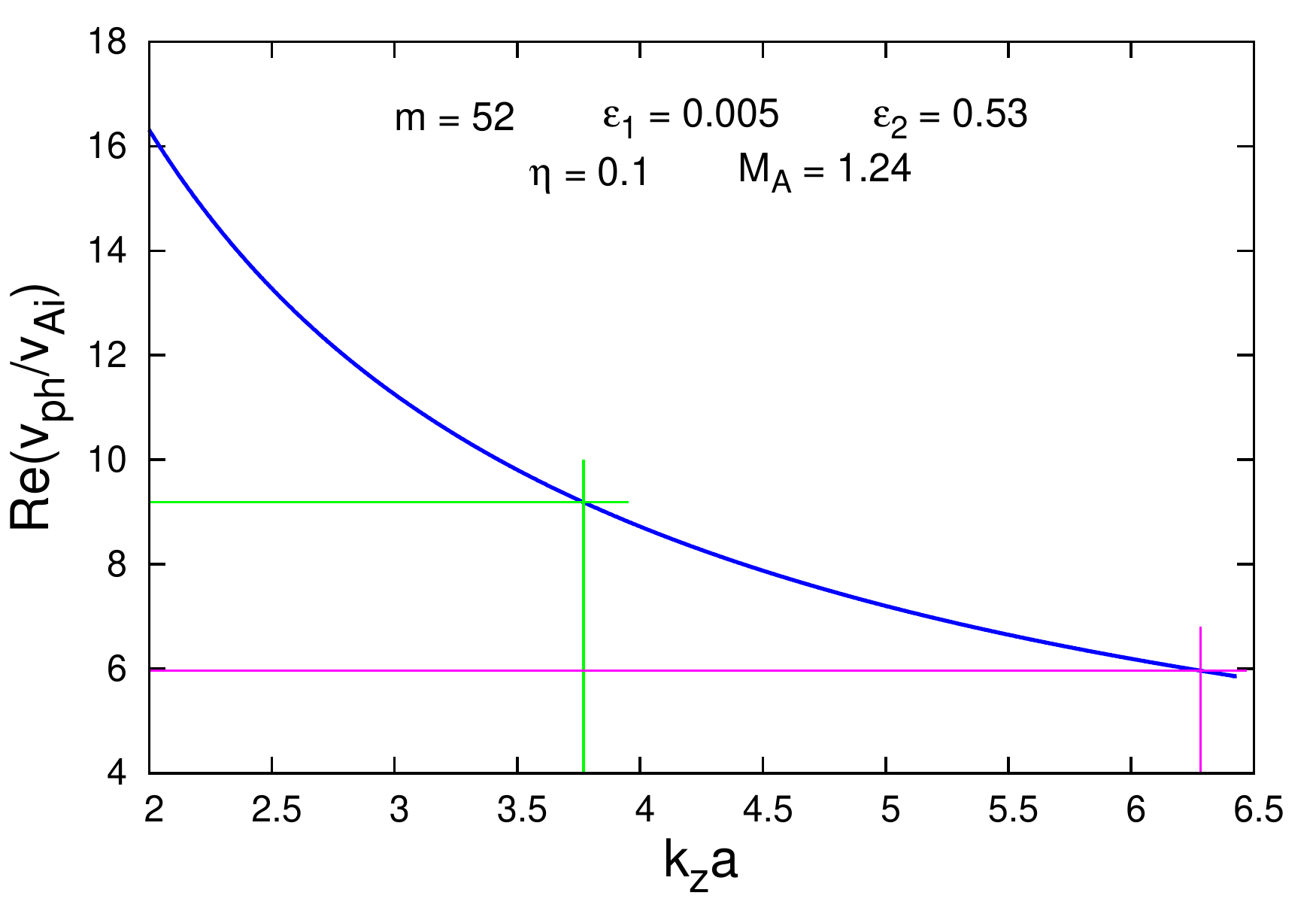}
               \hspace*{-0.03\textwidth}
               \includegraphics[width=0.515\textwidth,clip=]{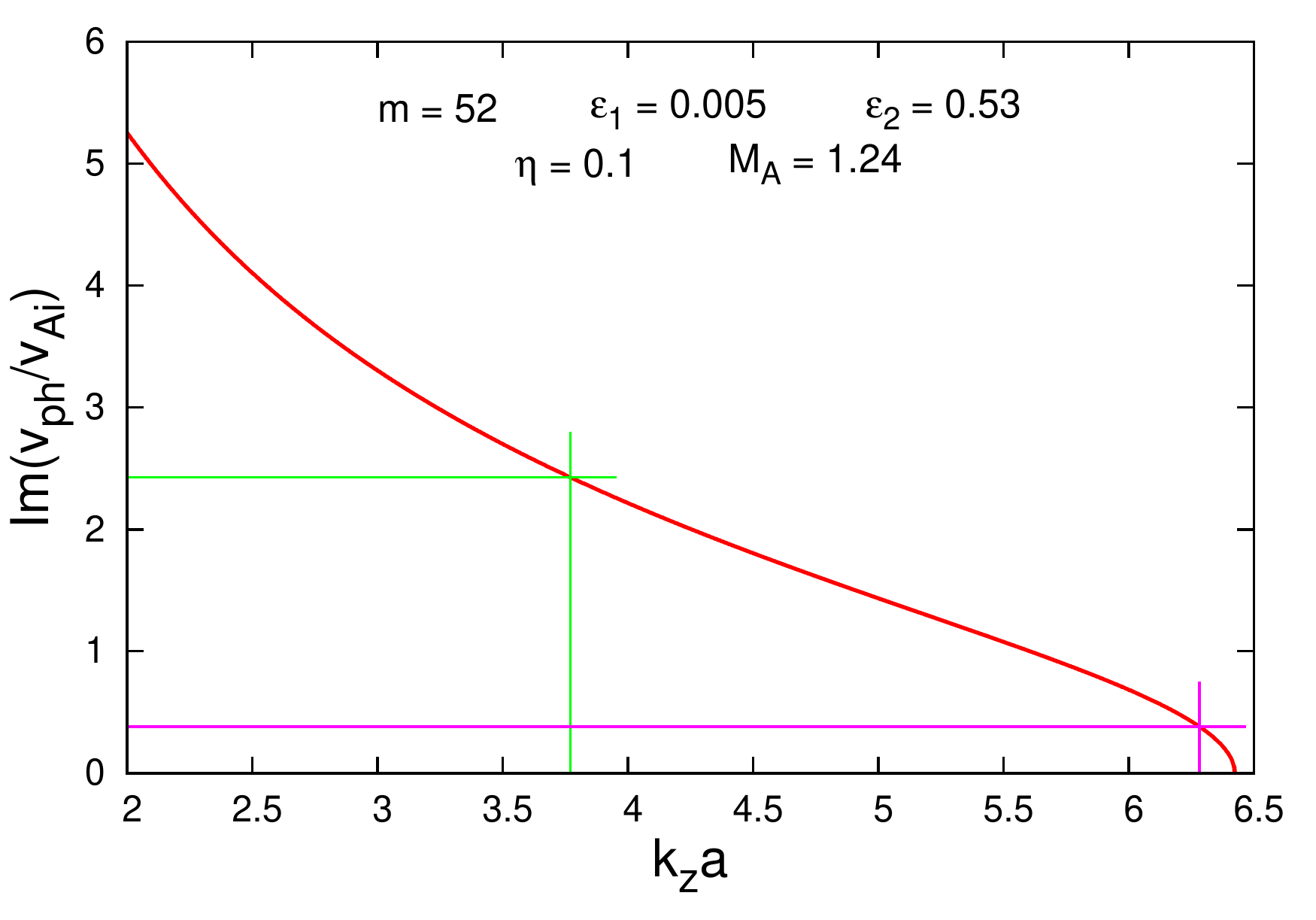}
              }
  \caption{(\emph{Left panel}) Dispersion curve of the $m = 52$ MHD mode propagating along a twisted incompressible macrospicule at $\eta = 0.1$, $b = 1.798$, $M_{\rm A} = 1.24$, $\varepsilon_1 = 0.005$, and $\varepsilon_2 = 0.53$.  (\emph{Right panel}) Normalized growth rate curve of the $m = 52$ MHD mode computed with the same input parameters as in the left panel.}
   \label{fig:fig1}
\end{figure}

Hence, the input parameters in the numerical task of solving Equation~\ref{eq:dispeq} are: $m = 52$, $\eta = 0.1$, $b = 1.798$, $\varepsilon_1 = 0.005$, $\varepsilon_2 = 0.53$, and $M_\mathrm{A} = 1.24$.  The results are pictured in Figure~\ref{fig:fig1}.
\begin{figure}[!ht]
   \centerline{\hspace*{0.015\textwidth}
               \includegraphics[width=0.515\textwidth,clip=]{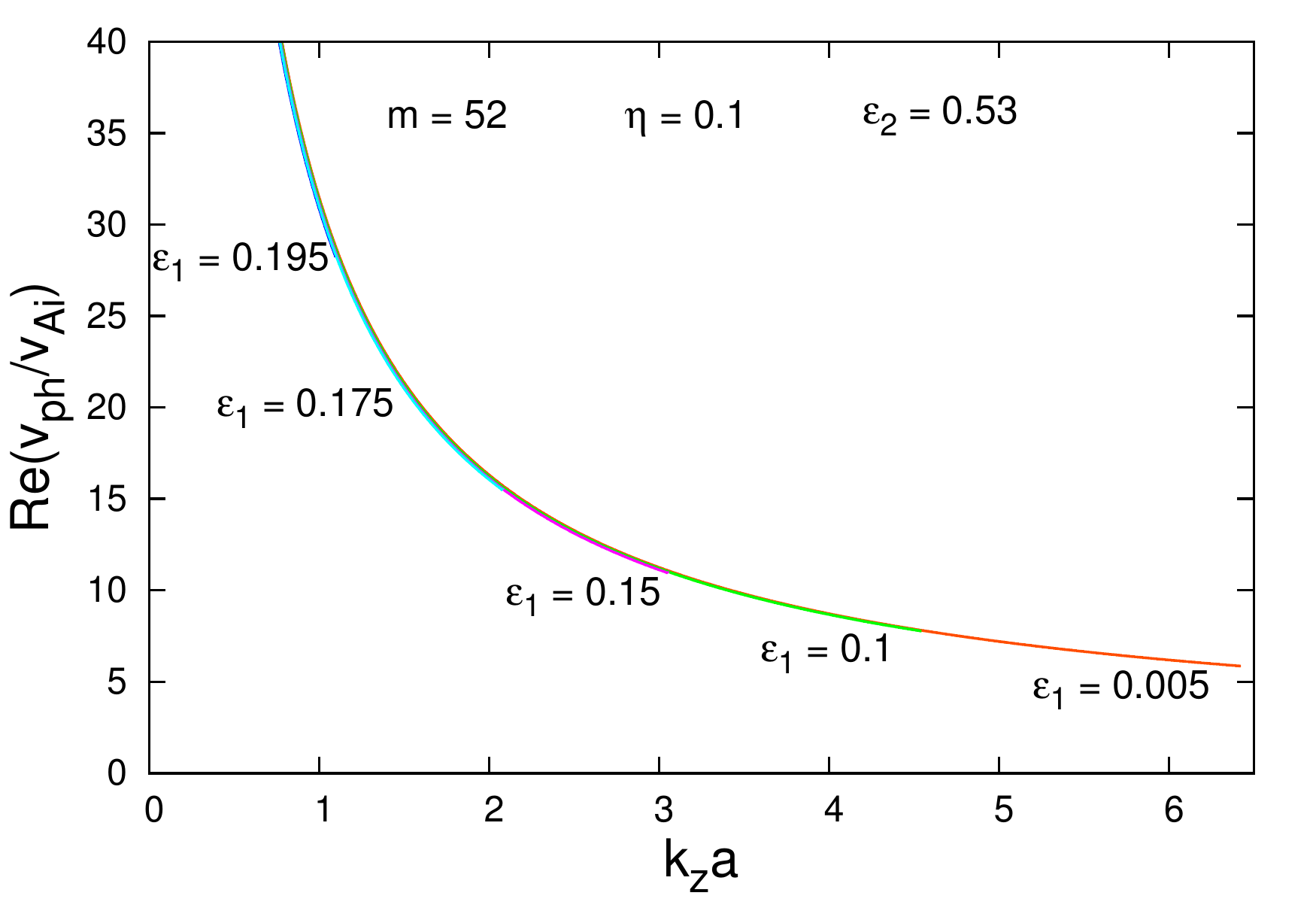}
               \hspace*{-0.03\textwidth}
               \includegraphics[width=0.515\textwidth,clip=]{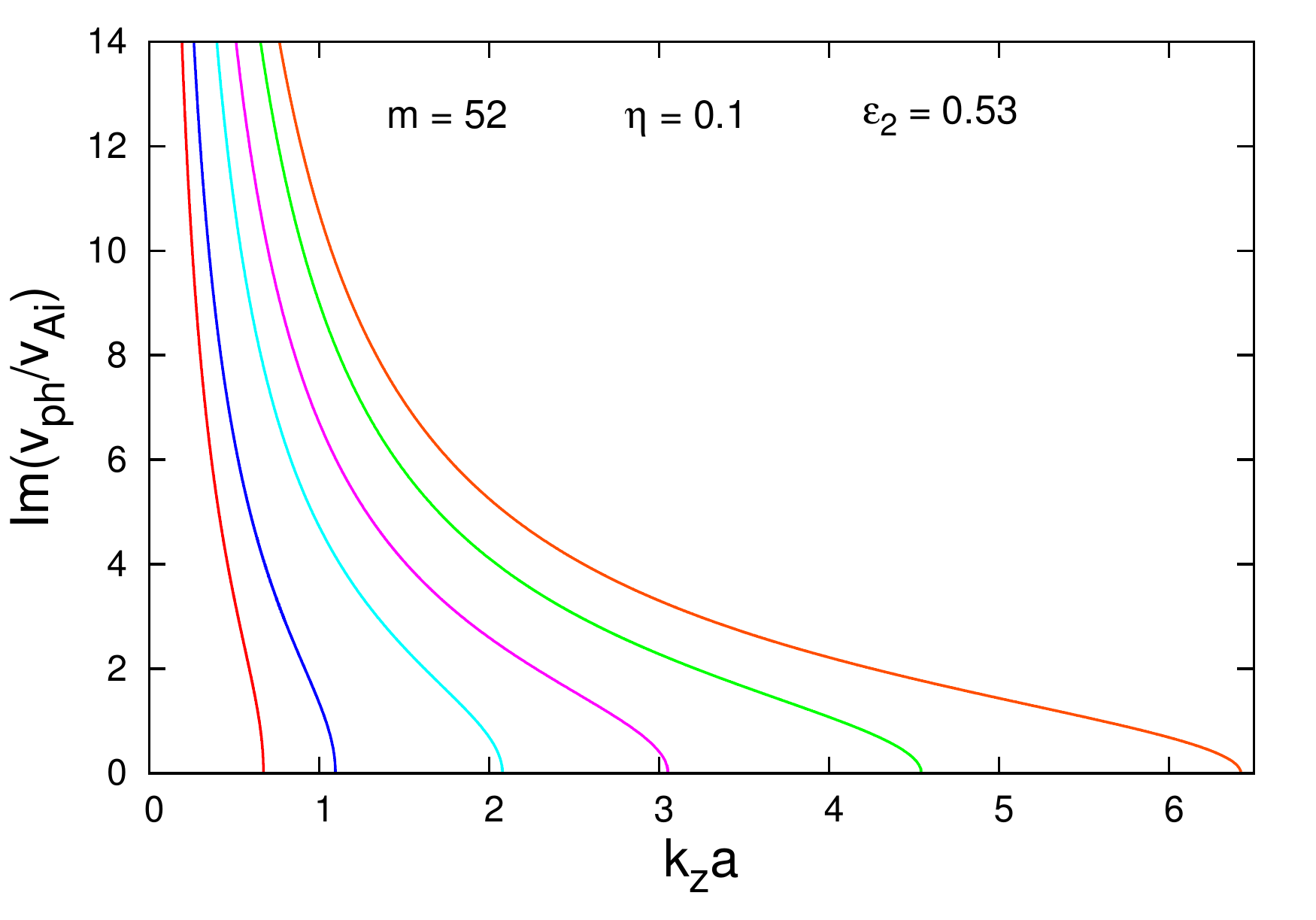}
              }
  \caption{(\emph{Left panel}) Dispersion curves of the unstable $m = 52$ MHD mode propagating along a twisted incompressible macrospicule at $\eta = 0.1$, $\varepsilon_2 = 0.53$, and the following values of $\varepsilon_1$ (from right to left): $0.005$, $0.1$, $0.15$, $0.175$, $0.195$, and $0.202085$ (red curve in the right plot).  Alfv\'en Mach numbers for these curves are respectively $1.24$, $1.23$, $1.22$, $1.22$, $1.21$, and $1.2119$.  (\emph{Right panel}) Growth rates of the unstable $m = 52$ mode for the same input parameters.  The azimuthal magnetic field that corresponds to $\varepsilon_1 = 0.202085$ (the instability window with zero width) and stops the KHI onset is equal to $0.57$~G.}
   \label{fig:fig2}
\end{figure}
From this plot we can calculate the instability characteristics of the $m = 52$ MHD mode for two unstable wavelengths, equal to $3$ and $5$~Mm ($k_z a = 6\pi/5$), respectively.  The KHI wave growth rate, $\gamma_\mathrm{KH}$, growth time, $\tau_\mathrm{KH} = 2\pi/\gamma_\mathrm{KH}$, as well as wave velocity, $v_\mathrm{ph}$, calculated from the graphics in Figure~\ref{fig:fig1}, for the aforementioned wavelengths are as follows:

For $\lambda_\mathrm{KH} = 3$~Mm we obtain
\[
    \gamma_\mathrm{KH} \cong 48.38 \times 10^{-3}\:\mathrm{s}^{-1}, \;\, \tau_\mathrm{KH} \cong 2.2\:\mathrm{min}, \;\, v_\mathrm{ph} \cong 361\:\mathrm{km}\,\mathrm{s}^{-1},
\]
while at $\lambda_\mathrm{KH} = 5$~Mm, we obtain
\[
    \gamma_\mathrm{KH} \cong 184.8 \times 10^{-3}\:\mathrm{s}^{-1}, \;\, \tau_\mathrm{KH} \cong 0.57\:\mathrm{min}, \;\, v_\mathrm{ph} \cong 556\:\mathrm{km}\,\mathrm{s}^{-1}.
\]
As seen, at both wavelengths phase velocities are super-Alfv\'enic.  The two growth times of $2.2$ and ${\approx}0.6$~min are reasonable taking into account that the macrospicule lifetime is about $15$~min, that is, the KHI at the selected wavelengths is rather fast.  A specific property of the instability $k_z a$ ranges is that for a fixed $m$, their widths depend upon the magnetic field twist parameter $\varepsilon_1$.  A discussion on this dependence is provided in \citet{zhelyazkov2018b} and here we quote it, namely ``with increasing the value of $\varepsilon_1$, the instability window becomes narrower and at some critical magnetic field twist its width equals zero.  This circumstance implies that for $\varepsilon_1 \geqslant \varepsilon_1^\mathrm{cr}$ there is no instability, or, in other words, there exists a critical azimuthal magnetic field $B_{\phi}^\mathrm{cr} = \varepsilon_1^\mathrm{cr} B_{\mathrm{i}z}$ that suppresses the instability onset''.  In Figure~\ref{fig:fig2}, a series of dispersion and dimensionless wave phase velocity growth rates for various increasing magnetic field twist parameter values has been plotted.  Note that each larger $\varepsilon_1$ implies an increase in $B_{\phi}$.  The red dispersion curve in the right panel of Figure~\ref{fig:fig2} has been obtained for $\varepsilon_1^\mathrm{cr} = 0.202085$ with $M_\mathrm{A} = 1.2119$, and it visually defines the left-hand-side limit of all other instability ranges.  The azimuthal magnetic field $B_{\phi}^\mathrm{cr}$ that stops the KHI is equal to $0.57$~G.

\begin{figure}[!ht]
   \centerline{\hspace*{0.015\textwidth}
               \includegraphics[width=0.515\textwidth,clip=]{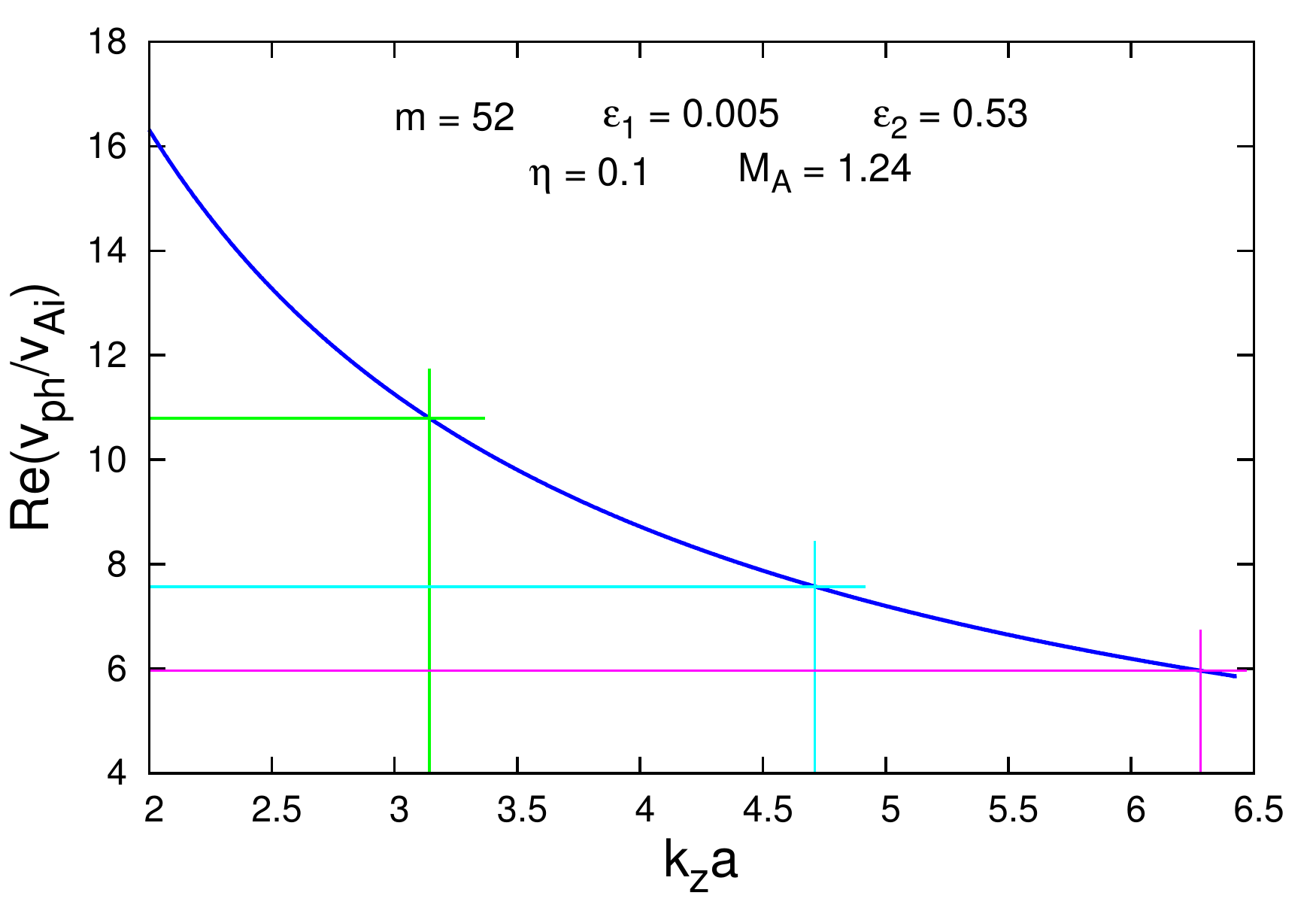}
               \hspace*{-0.03\textwidth}
               \includegraphics[width=0.515\textwidth,clip=]{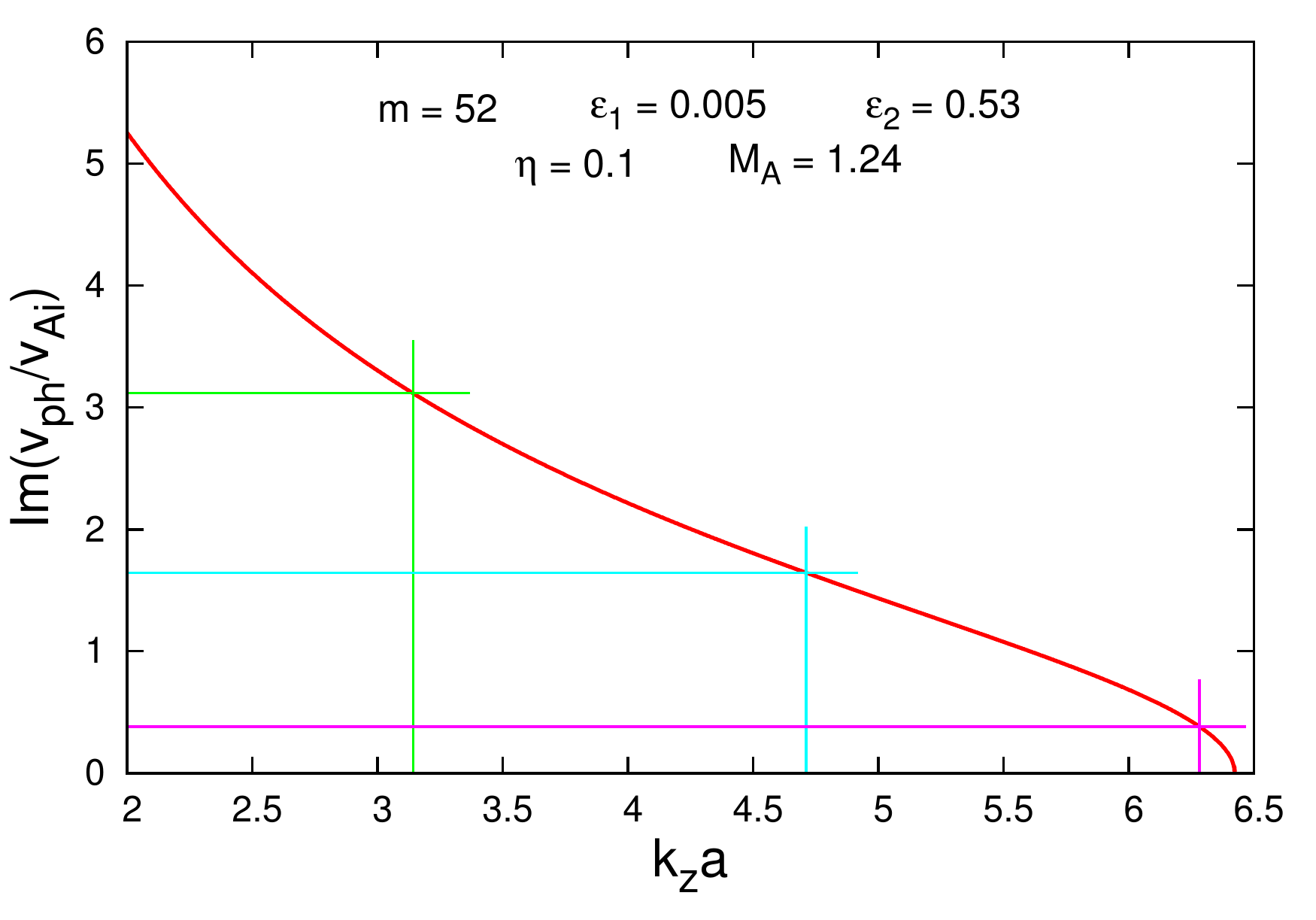}
              }
  \caption{(\emph{Left panel}) Dispersion curve of the $m = 52$ MHD mode propagating along a twisted incompressible macrospicule at $\eta = 0.1$, $b = 1.798$, $M_{\rm A} = 1.24$, $\varepsilon_1 = 0.005$, and $\varepsilon_2 = 0.53$.  The $k_z a$-positions marked with purple, cyan, and green vertical lines correspond to $\lambda_\mathrm{KH} = 4$~Mm for three different macrospicule widths, equal to $8$, $6$, and $4$~Mm, respectively.  (\emph{Right panel}) Normalized growth rate curve of the $m = 52$ MHD mode propagating along a twisted incompressible macrospicule with the same input parameters as in the left panel.}
   \label{fig:fig3}
\end{figure}

The instability $k_z a$-range of the $m = 52$ MHD mode pictured in Figure~\ref{fig:fig1} allows us to investigate how the width of the macrospicule will affect the KHI characteristics for a fixed instability wavelength.  Such an appropriate wavelength is $\lambda_\mathrm{KH} = 4$~Mm.  We calculate (and plot) the instability growth rate, $\gamma_\mathrm{KH}$, the instability development or growth time, $\tau_\mathrm{KH}$, and the wave phase velocity of the $m = 52$ mode.  Our choice for macrospicule widths is: $8$, $6$, and $4$~Mm, respectively.  Note that the unstable $4$~Mm wavelength has three different positions on the $k_z a$-axis, notably $k_z a = 2\pi$ for $\Delta \ell = 8$~Mm, $k_z a = 1.5\pi$ for $\Delta \ell = 6$~Mm, and $k_z a = \pi$ for $\Delta \ell = 4$~Mm (see Figure~\ref{fig:fig3}).  The basic KHI characteristics of the $m = 52$ MHD mode at
\begin{table}
\caption{Kelvin--Helmholtz instability characteristics of the $m = 52$ MHD mode at $\lambda_\mathrm{KH} = 4$~Mm for three different widths of the macrospicule correspondingly equal to $8$, $6$, and $4$~Mm.}
\label{tab:khiparameters}
\vspace*{2mm}


\begin{tabular}{cccccc}
\hline
$\Delta \ell$ & $\gamma_\mathrm{KH}$ & $\tau_\mathrm{KH}$ & $v_\mathrm{ph}$ & $\varepsilon_1^{\mathrm{cr}}$ & $B_\phi^{\mathrm{cr}}$\\
 (Mm) & (${\times}10^{-3}$ s$^{-1}$) &  (min) & (km\,s$^{-1}$) &   &  (G) \\
\hline
8 &  36.28 & 2.9  & 361 & 0.19843   & 0.55  \\
6 & 156.48 & 0.67 & 459 & 0.202085  & 0.57  \\
4 & 246.70 & 0.35 & 654 & 0.205465  & 0.58  \\
\hline
\end{tabular}
\end{table}
the wavelength $\gamma_\mathrm{KH} = 4$~Mm at the aforementioned three different macrospicule widths are presented in Table~\ref{tab:khiparameters}.  The most striking result concerns the instability growth or developing time: it is only approximately half a minute at $\Delta \ell = 4$~Mm and approximately seven times longer ($2.9$ min) when the jet width is $8$~Mm.  It is not surprising, bearing in mind the shape of the dispersion curve of the $m = 52$ MHD mode, that the wave phase velocities of the unstable mode quickly increase from $361$~km\,s$^{-1}$ at the widest macrospicule to $654$~km\,s$^{-1}$ at the narrowest one.  In that table, we also give the critical magnetic field twist parameter, $\varepsilon_1^{\mathrm{cr}}$, at which the size of the instability range of a given jet becomes equal to zero.  Those three limiting wave growth rate curves are plotted in
\begin{figure}[!ht]
   \centerline{\hspace*{0.015\textwidth}
               \includegraphics[width=0.7\textwidth,clip=]{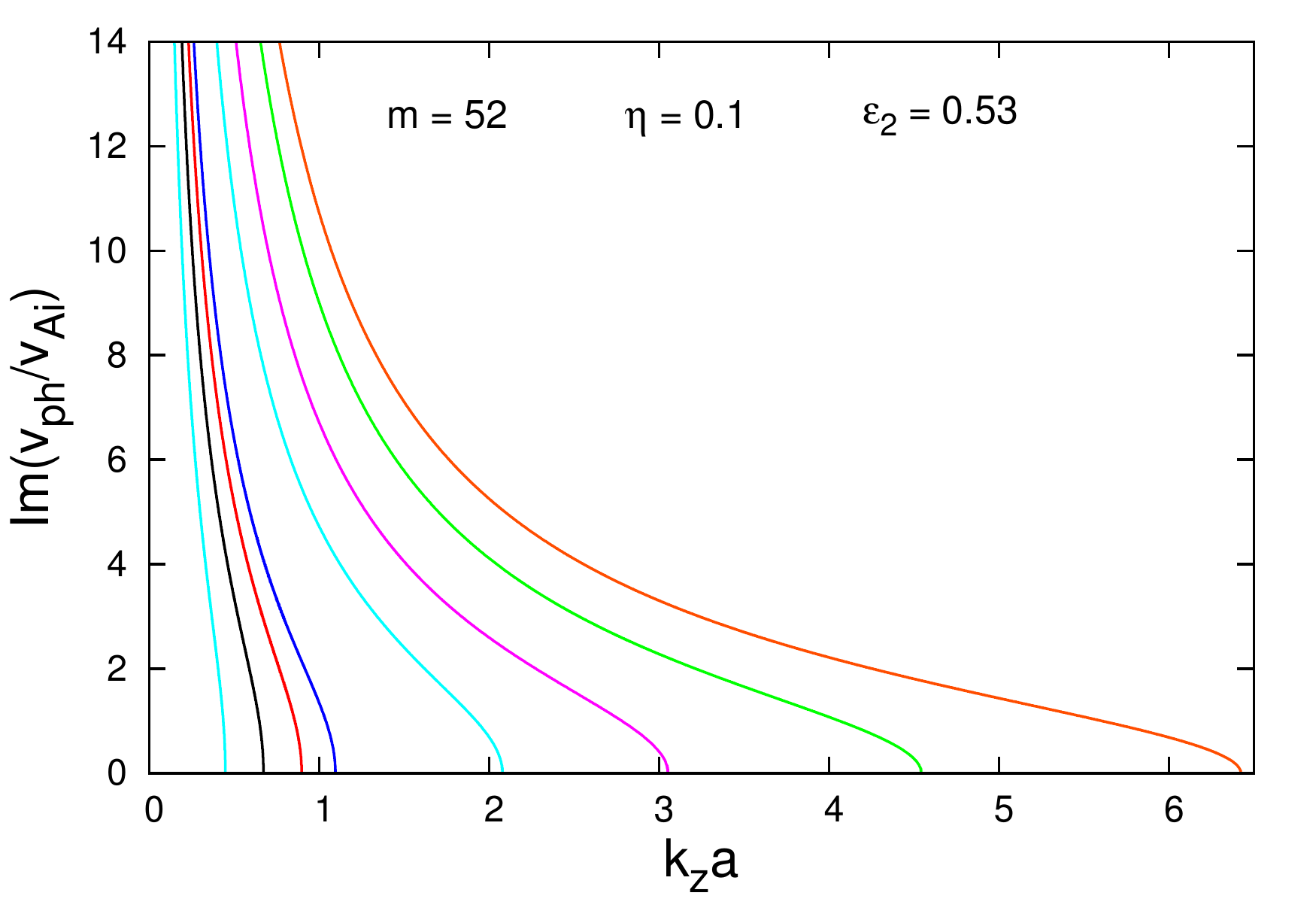}
              }
  \caption{Growth rate curves of the $m = 52$ MHD mode propagating along a twisted incompressible macrospicule at $\eta = 0.1$, $\varepsilon_2 = 0.53$, and 8 different values of $\varepsilon_1$ equal to $0.005$ (orange curve), $0.1$ (green curve), $0.15$ (purple curve), $0.175$ (cyan curve), $0.195$ (blue curve), $0.19843$ (red curve), $0.202085$ (black curve), and $0.205465$ (cyan curve), respectively.  The growth rate curves computed with the last three values of the magnetic field twist parameter $\varepsilon_1$ correspond to the three different macrospicule widths, equal to $8$, $6$, and $4$~Mm, respectively.  The Alfv\'en Mach numbers used at the computation of these three curves are accordingly equal to $1.2128$, $1.2119$, and $1.211$.}
   \label{fig:fig4}
\end{figure}
three different colors in Figure~\ref{fig:fig4}; from left to right the cyan curve corresponds to $\Delta \ell = 4$~Mm, the black one to $6$~Mm, and the red curve to the macrospicule width of $8$~Mm.  It is interesting to observe that the critical azimuthal magnetic field components, $B_\phi^{\mathrm{cr}}$, that would stop the KHI appearance have very close magnitudes, roughly $0.6$~G.  The plots of the dimensionless dispersion curves for the last three values of $\varepsilon_1$ in Figure~\ref{fig:fig4}, similar to those seen in the left panel of Figure~\ref{fig:fig2}, are practically not usable---they possess very high values (in the range of $80$--$140$) corresponding to extremely high wave phase velocities.

Another interesting observation is that the KHI growth or developing times of the $m = 52$ MHD mode, evaluated at wavelengths equal to the half-width of the macro\-spicule and, say at $(\Delta \ell/2 + 2$)~Mm, for the three aforementioned widths are practically on the same order.  This observation is illustrated in Table~\ref{tab:growthtimes}.  At the thinnest jet both growth times are the shortest while at the thickest jet they are relatively longer.  An observational evaluation of the macrospicule width will naturally help in finding the appropriate conditions for KHI onset of the corresponding high MHD mode.
\begin{table}
\caption{Kelvin--Helmholtz instability growth times of the $m = 52$ MHD mode at wavelengths equal to the half-width of the jet and at $(\Delta \ell/2 + 2)$~Mm for three different widths of the macrospicule correspondingly equal to $8$, $6$, and $4$~Mm.}
\label{tab:growthtimes}
\vspace*{2mm}


\begin{tabular}{ccc}
\hline
$\Delta \ell$ & $\tau_\mathrm{KH}(\Delta \ell/2)$ & $\tau_\mathrm{KH}[(\Delta \ell/2 + 2)$~Mm]\\
 (Mm) & (min) &  (min) \\
\hline
8 &  2.9  &  0.80  \\
6 &  2.2  &  0.57  \\
4 &  1.4  &  0.35  \\
\hline
\end{tabular}
\end{table}

It is interesting to see how the change in background magnetic field, $B_\mathrm{e}$, will modify the picture.  We set $B_\mathrm{e} = 4.8$~G (a magnetic field at which the total balance Equation~\ref{eq:pbeq} is satisfied)---with the same input parameters for electron number densities and electron temperatures, we have new values for Alfv\'en speeds, plasma betas, and magnetic fields ratio, notably $v_\mathrm{Ai} = 52.36$~km\,s$^{-1}$, $v_\mathrm{Ae} = 330.9$~km\,s$^{-1}$, $\beta_\mathrm{i} = 3.01$, $\beta_\mathrm{e} = 0.15$, and $b = 1.998$.  With this new internal Alfv\'en speed, $v_\mathrm{Ai}$, the Alfv\'en Mach number has a slightly higher magnitude, $M_\mathrm{A} = 1.43$.  It turns out that the appropriate MHD mode number that provides the required width of the instability window at $\Delta \ell = 6$~Mm is $m = 48$.  Performing the calculations with $m = 48$, $\eta = 0.1$, $\varepsilon_1 = 0.005$, $\varepsilon_2 = 0.53$, $b = 1.998$, and $M_\mathrm{A} = 1.43$, we obtain dispersion and growth rate curves that are very similar to those pictured in Figure~\ref{fig:fig1}.  We do not plot these curves, but compare the instability characteristics for both mode numbers, $48$ and $52$, respectively, at the instability wavelength of $3$ and $5$~Mm.  The results of this comparison are shown in Table~\ref{tab:comparison}.
\begin{table}
\caption{Kelvin--Helmholtz instability characteristics of the $m = 48$ and $52$ MHD modes at $\lambda_\mathrm{KH} = 3$ and $5$~Mm for the macrospicule width equal to $6$~Mm.}
\label{tab:comparison}
\vspace*{2mm}


\begin{tabular}{cccccc}
\hline
Mode number & $\gamma_\mathrm{KH}$ & $\tau_\mathrm{KH}$ & $v_\mathrm{ph}$ & $\varepsilon_1^{\mathrm{cr}}$ & $B_\phi^{\mathrm{cr}}$\\
 $m$ & (${\times}10^{-3}$ s$^{-1}$) &  (min) & (km\,s$^{-1}$) &   &  (G) \\
\hline
    &   & $\lambda_{\mathrm{KH}} = 3$~Mm &     & \\
\hline
48 &  49.23 & 2.1  & 338.0 & 0.23494    & 0.58  \\
52 &  48.38 & 2.2  & 361.0 & 0.202085   & 0.57  \\
\hline
    &   & $\lambda_{\mathrm{KH}} = 5$~Mm &     & \\
\hline
48 &  172.96 & 0.61  & 518.0 & 0.23494    & 0.48  \\
52 &  184.85 & 0.57  & 556.5 & 0.202085   & 0.58  \\
\hline
\end{tabular}
\end{table}

As seen, the KHI characteristics for the two mode numbers of $48$ and $52$ are very close.  One can claim that at each MHD mode number that ensures an instability window width like the one shown in Figure~\ref{fig:fig1} will yield similar KHI characteristics as those in Table~\ref{tab:comparison}.

\section{Summary and Conclusion}
\label{sec:conclusion}
In this article, we have established the conditions under which high MHD modes excited in a solar macrospicule can become unstable against the Kelvin--Helmholtz instability.  We modeled the jet as an axially moving, weakly twisted cylindrical magnetic flux tube of radius $a$ and homogeneous density $\rho_\mathrm{i}$ that rotates around its axis, surrounded by coronal plasma with homogeneous magnetic field $B_\mathrm{e}$ and density $\rho_\mathrm{e}$.  The twist of the internal magnetic field $\vec{B}_\mathrm{i}$ is characterized by the ratio $B_{\mathrm{i}\phi}/B_{\mathrm{i}z} = \varepsilon_1$, where the components of the internal magnetic field, $\vec{B}_\mathrm{i}$, are evaluated at the tube radius, $a$.  In a similar way, the macrospicule velocity twist is specified by the ratio $U_\phi/U_z$, where $U_\phi$ and $U_z$ are the rotational and axial speeds of the jet.  Along with these two parameters, the density contrast $\rho_\mathrm{e}/\rho_\mathrm{i} = \eta$ plays an important role in the modeling.  Our choice of that parameter is $0.1$ because we assume that the macrospicule should be at least one order denser that its environment.  Other important physical parameters are the plasma betas of both media---their values tell us how to treat each medium, as incompressible or cool plasma---the general case of compressible media is still intractable from a theoretical point of view.  Electron temperatures and background or jet magnetic field, at the given density contrast, actually control the values of plasma betas.  With $T_\mathrm{e} = 1.0$~MK, $T_\mathrm{i} = 500\,000$~K, and $B_\mathrm{e} = 5$~G, as well as $U_\phi = 40$~km\,s$^{-1}$ and $\varepsilon_1 = 0.005$, from the total pressure balance Equation \ref{eq:pbeq} we obtain $\beta_\mathrm{i} \cong 2.25$ and $\beta_\mathrm{e} \cong 0.14$.  These plasma beta values imply that the macrospicule medium can be treated as incompressible plasma while the surrounding magnetized plasma may be considered a cool medium \citep{zank1993}.  It is worth noting, however, that a decrease in the macrospicule temperature to $300\,000$~K will diminish $\beta_\mathrm{i}$ to $0.71$, thus making the jet plasma treatment as incompressible medium problematic.  When the two media are treated as cool magnetized plasmas an acceptable modeling of the KHI in our macrospicule would require the excitation of a mode higher than the $m = 52$ MHD mode.

The dispersion Equation \ref{eq:dispeq} yields unstable solutions for each mode number $m \geqslant 2$.  For relatively small MHD mode numbers, however, the shortest unstable wavelengths that can be `extracted' at $k_z a$-positions near the right-hand-side limit in Equation~\ref{eq:instcond} of the instability window are too long to be comparable with the sizes of KH vortex-like blobs appearing at the macrospicule interface.  Reliable unstable wavelengths are achieved at the excitation of very high MHD modes.  This is not surprising for chromospheric--TR jets: for instance, \citet{kuridze2016} investigated the dynamics and stability of small-scale rapid redshifted and blueshifted excursions, that appear as high-speed jets in the wings of the H$\alpha$ line, explained their short lifetimes (a few seconds) as a result of a KHI that arises in excited high-mode MHD waves.  To achieve growth times of a few seconds using a dispersion equation similar to Equation~\ref{eq:dispeq} (but with $\kappa_\mathrm{e} = k_z$) it was necessary to assume azimuthal mode numbers up to $100$.  In our case an $m = 52$ makes the instability region wide enough to accommodate unstable wavelengths of at least $3$~Mm and a KHI growth time of $2.2$~min.  This growth time becomes shorter when the wavelength increases; for example at $\gamma_\mathrm{KH} = 4$~Mm the growth time is around $0.7$~min, while at $\gamma_\mathrm{KH} = 5$~Mm it is equal to ${\cong}0.6$~min.  We have also studied how small variations of the macrospicule width affect the instability development time of the excited MHD wave with mode number $m = 52$, a decrease in $\Delta \ell$ to $4$~Mm yields generally shorter growth times, while an increase of macrospicule width to $8$~Mm gives longer instability growth times.  Except through the change of the azimuthal mode number $m$, the width of the instability window can be regulated by increasing or decreasing the magnetic field twist parameter $\varepsilon_1$.  A progressive increase of the magnetic field twist parameter, $\varepsilon_1$, yields a further decrease of the instability region width and at some critical $\varepsilon_1$, the width becomes equal to zero, that is, there is no instability at all.  This $\varepsilon_1^{\mathrm{cr}}$ defines an azimuthal internal magnetic field component, $B_{\mathrm{i}\phi}^\mathrm{cr}$, that stops the KHI appearance.  For our macrospicule this critical magnetic field is relatively small, it is equal to $0.6$~G.  Such a small azimuthal field component of the twisted magnetic field might be a reason for the inability to easily observe or detect KH features in spinning macrospicules.  The shape of the dimensionless wave dispersion curve, shown in the left panels of Figures~\ref{fig:fig1} and \ref{fig:fig3}, tells us that the excited $m = 52$ MHD mode is a super-Alfv\'enic wave whose phase velocity grows very fast with the the increase of the wavelength.  A realistic modeling of unstable MHD modes therefore requires the finding of such an azimuthal mode number, $m$, which will ensure an instability region with a width that should contain the expected unstable wavelengths near to its right-hand-side limit $(k_z a)_\mathrm{rhs}$.  A change in the background magnetic field, $B_\mathrm{e}$, can influence the MHD mode number $m$, which would yield an instability region similar or identical to that seen in Figures~\ref{fig:fig1} and \ref{fig:fig3}.  For instance, at a weaker environment magnetic field $B_\mathrm{e} = 4.8$~G an almost identical instability window can be achieved with the excitation of the $m = 48$ MHD wave.  The KHI characteristics at this mode number are very close to those obtained at the excitation of the $52$ mode number. A comparison of this is shown in Table~\ref{tab:comparison}.

The results obtained in this article can be influenced by assuming more complicated velocity and magnetic field profiles, as well as radially inhomogeneous plasma densities.  The latter will involve the appearance of continuous spectra and resonant wave absorption, which will modify the KHI characteristics to some extent.  The nonlinearity leads to the saturation of the KHI growth and to the formation of nonlinear waves \citep{miura1984}.  Our approach, is nonetheless flexible enough and can yield reasonable growth times of observationally detected Kelvin--Helmholtz instabilities in solar atmospheric jets.  The next step of its improvement is to include compressibility in governing MHD equations.  Arising KHI in the small-scale chromospheric jets, such as macrospicules, and the triggered wave turbulence can contribute to coronal heating and to the energy balance in the solar transition region.

\section*{Appendix A: Derivation of the Wave Dispersion Relation}

We recall that we modeled the spinning macrospicule as a rotating and axially moving twisted magnetic flux tube of radius $r = a$.  In a cylindrical coordinate system the magnetic and velocity fields inside the jet are assumed to be
\[
    \left( 0, B_{\mathrm{i}\phi}(r), B_{\mathrm{i}z} \right) \quad \mbox{and} \quad \left( 0, U_\phi(r), U_{z} \right),
\]
respectively.  Linearized ideal MHD equations, which govern the incompressible dynamics of perturbations in the rotating jet are
\begin{equation}
\label{eq:momentum}
    \frac{\partial}{\partial t}\boldsymbol{v} + (\boldsymbol{U}\cdot \nabla)\boldsymbol{v} + (\boldsymbol{v}\cdot \nabla) \boldsymbol{U} = -\frac{\nabla p_\mathrm{tot}}{\rho_\mathrm{i}} + \frac{\left( \boldsymbol{B}_\mathrm{i} \cdot \nabla \right)\boldsymbol{b}}{\rho_\mathrm{i} \mu} + \frac{\left( \boldsymbol{b} \cdot \nabla \right)\boldsymbol{B}_\mathrm{i}}{\rho_\mathrm{i} \mu},
\end{equation}
\begin{equation}
\label{eq:induct}
	\frac{\partial}{\partial t}\boldsymbol{b} - \nabla \times \left( \boldsymbol{v}
    \times \boldsymbol{B}_\mathrm{i} \right) - \nabla \times \left( \boldsymbol{U} \times \boldsymbol{b}
    \right) = 0,	
\end{equation}
\begin{equation}
\label{eq:divv}
	\nabla \cdot \boldsymbol{v} = 0,	
\end{equation}
\begin{equation}
\label{eq:divb}
	\nabla \cdot \boldsymbol{b} = 0,
\end{equation}
where $\boldsymbol{v} = (v_r, v_\phi, v_z)$ and $\boldsymbol{b} = (b_r, b_\phi, b_z)$ are the perturbations of fluid velocity and magnetic field, respectively, and $p_\mathrm{tot}$ is the perturbation of the total pressure $p_\mathrm{t}$.

Assuming that all perturbations are ${\propto}g(r)\exp \left[\mathrm{i} \left( -\omega t + m \phi + k_z z \right) \right]$, with $g(r)$ being just a function of $r$, we obtain from the above set of equations the following ones:
\begin{equation}
\label{eq:moment-r}
    -\mathrm{i}\sigma v_r -2\frac{U_\phi}{r}v_\phi - \mathrm{i}\frac{f_B}{\mu \rho_\mathrm{i}}b_r + 2\frac{B_{\mathrm{i}\phi}}{\mu \rho_\mathrm{i} r}b_\phi = -\mathrm{i}\frac{1}{\rho_\mathrm{i}}\frac{\mathrm{d}p_\mathrm{tot}}{\mathrm{d}r},
\end{equation}
\begin{equation}
\label{eq:moment-phi}
    -\mathrm{i}\sigma v_\phi + \frac{1}{r}\frac{\mathrm{d}\left( r U_\phi \right)}{\mathrm{d}r}v_r - \mathrm{i}\frac{f_B}{\mu \rho_\mathrm{i}}b_\phi
    - \frac{1}{\mu \rho_\mathrm{i}} \frac{1}{r}\frac{\mathrm{d}\left( r B_{\mathrm{i}\phi} \right)}{\mathrm{d}r}b_r = -\mathrm{i} \frac{1}{\rho_\mathrm{i}}\frac{m}{r}p_\mathrm{tot},
\end{equation}
\begin{equation}
\label{eq:moment-z}
    -\mathrm{i}\sigma v_z - \mathrm{i}\frac{f_B}{\mu \rho_\mathrm{i}}b_z = -\mathrm{i}\frac{1}{\rho_\mathrm{i}}k_z p_\mathrm{tot},
\end{equation}
\begin{equation}
\label{eq:induct-r}
    -\mathrm{i}\sigma b_r - \mathrm{i}f_B v_r = 0,
\end{equation}
\begin{equation}
\label{eq:induct-phi}
    -\mathrm{i}\sigma b_\phi - r\frac{\mathrm{d}}{\mathrm{d}r}\left( \frac{U_\phi}{r} \right)b_r - \mathrm{i}f_B v_\phi + r\frac{\mathrm{d}}{\mathrm{d}r}\left( \frac{B_{\mathrm{i}\phi}}{r} \right)v_r = 0,
\end{equation}
\begin{equation}
\label{eq:induct-z}
    -\mathrm{i}\sigma b_z - \mathrm{i}f_B v_z = 0,
\end{equation}
\begin{equation}
\label{eq:nablav}
    \left( \frac{\mathrm{d}}{\mathrm{d}r} + \frac{1}{r} \right)v_r + \mathrm{i}\frac{m}{r}v_\phi + \mathrm{i}k_z v_z = 0,
\end{equation}
where
\begin{equation}
\label{eq:sigma}
    \sigma = \omega - \frac{m}{r}U_\phi - k_z U_z
\end{equation}
is the Doppler-shifted frequency and
\begin{equation}
\label{eq:fB}
    f_B = \frac{m}{r}B_{\mathrm{i}\phi} + k_z B_{\mathrm{i}z}.
\end{equation}

It is now convenient to introduce the Lagrangian displacement, $\bm{\xi}$, and express it via the fluid velocity perturbation through the relation \citep{chandrasekhar1961}
\[
    \boldsymbol{v} = \frac{\partial \bm{\xi}}{\partial t} + \left( \boldsymbol{U}\cdot \nabla \right)\bm{\xi} - \left( \boldsymbol{\bm{\xi}}\cdot \nabla \right)\boldsymbol{U},
\]
which yields
\begin{equation}
\label{eq:xi}
    v_r = -\mathrm{i}\sigma \xi_r, \qquad v_\phi = -\mathrm{i}\sigma \xi_\phi - r\frac{\mathrm{d}}{\mathrm{d}r}\left( \frac{U_\phi}{r} \right)\xi_r, \qquad v_z = -\mathrm{i}\sigma \xi_z.
\end{equation}

In terms of $\bm{\xi}$, Equations~\ref{eq:moment-r}--\ref{eq:nablav} can be rewritten as
\begin{equation}
\label{eq:xi1}
    \left[ \sigma^2 - \omega_\mathrm{Ai}^2 - r\frac{\mathrm{d}}{\mathrm{d}r} \left( \frac{U_\phi^2}{r^2} \right) + \frac{1}{\mu \rho_\mathrm{i}}r\frac{\mathrm{d}}{\mathrm{d}r} \left( \frac{B_{\mathrm{i}\phi}^2}{r^2} \right) \right]\xi_r
    - 2\mathrm{i}\left( \sigma \frac{U_\phi}{r} + \frac{1}{r}\frac{B_{\mathrm{i}\phi}f_B}{\mu \rho_\mathrm{i}} \right)\xi_\phi = \frac{1}{\rho_\mathrm{i}}\frac{\mathrm{d}p_\mathrm{tot}}{\mathrm{d}r},
\end{equation}
\begin{equation}
\label{eq:xi1}
    \left( \sigma^2 - \omega_\mathrm{Ai}^2 \right)\xi_\phi + 2\mathrm{i}\left( \sigma \frac{U_\phi}{r} + \frac{1}{r}\frac{B_{\mathrm{i}\phi}f_B}{\mu \rho_\mathrm{i}} \right)\xi_r = \mathrm{i} \frac{1}{\rho_\mathrm{i}}\frac{m}{r}p_\mathrm{tot},
\end{equation}
\begin{equation}
\label{eq:xi2}
    \left( \sigma^2 - \omega_\mathrm{Ai}^2 \right)\xi_z = \mathrm{i} \frac{1}{\rho_\mathrm{i}}k_z p_\mathrm{tot},
\end{equation}
\begin{equation}
\label{eq:xi3}
    \left( \frac{\mathrm{d}}{\mathrm{d}r} + \frac{1}{r} \right)\xi_r + \mathrm{i}\frac{m}{r}\xi_\phi + \mathrm{i}k_z \xi_z = 0,
\end{equation}
where
\begin{equation}
\label{eq:omegaAi}
    \omega_\mathrm{Ai} = \frac{f_B}{\sqrt{\mu \rho_\mathrm{i}}}
\end{equation}
is the local Alfv\'en frequency inside the jet.

Excluding $\xi_\phi$ and $\xi_z$ from these equations, we obtain
\begin{equation}
\label{eq:xi_r}
    \rho_\mathrm{i} \left( \sigma^2 - \omega_\mathrm{Ai}^2 \right)\left( \frac{\mathrm{d}\xi_r}{\mathrm{d}r} + \frac{\xi_r}{r} \right) + 2\rho_\mathrm{i}d_2 \frac{m}{r}\xi_r = \left( \frac{m^2}{r^2} + k_z^2 \right)p_\mathrm{tot},
\end{equation}
\begin{equation}
\label{eq:xi_r-new}
    \rho_\mathrm{i}d_1 \xi_r = \left( \sigma^2 - \omega_\mathrm{Ai}^2 \right)\frac{\mathrm{d}p_\mathrm{tot}}{\mathrm{d}r} - 2\frac{m}{r}d_2 p_\mathrm{tot},
\end{equation}
where
\[
    d_1 = \left( \sigma^2 - \omega_\mathrm{Ai}^2 \right)^2 - r\left( \sigma^2 - \omega_\mathrm{Ai}^2 \right) \left[ \frac{\mathrm{d}}{\mathrm{d}r}\left( \frac{U_{\phi}^2}{r^2} \right) - \frac{1}{\mu \rho_\mathrm{i}}\frac{\mathrm{d}}{\mathrm{d}r}\left( \frac{B_{\mathrm{i}\phi}^2}{r^2} \right) \right] - 4d_2^2,  \;
    d_2 = \sigma\frac{U_{\phi}}{r} + \frac{B_{\mathrm{i}\phi} f_B}{\mu \rho_\mathrm{i}r}.
\]
By presenting $\xi_r$ from Equation~\ref{eq:xi_r-new} in terms of $p_\mathrm{tot}$ and inserting it into Equation~\ref{eq:xi_r}, we obtain the following equation for the total pressure perturbation:
\begin{eqnarray}
\label{eq:ptoteq}
    \left[ \left( \sigma^2 - \omega_\mathrm{Ai}^2 \right)\frac{\mathrm{d}}{\mathrm{d}r} + \frac{\sigma^2 - \omega_\mathrm{Ai}^2 + 2m d_2}{r} \right] \nonumber \\
    \nonumber \\
    {}\times\left[ \frac{\sigma^2 - \omega_\mathrm{Ai}^2}{d_1}\frac{\mathrm{d}p_\mathrm{tot}}{\mathrm{d}r} - \frac{2md_2}{d_1}\frac{p_\mathrm{tot}}{r} \right] - \left( \frac{m^2}{r^2} + k_z^2 \right)p_\mathrm{tot} = 0.
\end{eqnarray}
This equation can be significantly simplified by considering that the rotation and the magnetic twist of the jet are uniform, that is,
\begin{equation}
\label{eq:uniform}
    U_\phi(r) = \Omega r \qquad  \mbox{and} \qquad B_\phi(r) = A r,
\end{equation}
where $\Omega$ and $A$ are constants.  In this case, Equation~\ref{eq:ptoteq} takes the form of the Bessel equation
\begin{equation}
\label{eq:bessel}
    \frac{\mathrm{d}^2p_\mathrm{tot}}{\mathrm{d}r^2} + \frac{1}{r}\frac{\mathrm{d}p_\mathrm{tot}}{\mathrm{d}r} - \left( \frac{m^2}{r^2} + \kappa_\mathrm{i}^2 \right)p_\mathrm{tot} = 0,
\end{equation}
where
\begin{equation}
\label{eq:kappa}
    \kappa_\mathrm{i}^2 = k_z^2\left[ 1 - 4\left( \frac{\sigma \Omega + A\omega_\mathrm{Ai}/\!\sqrt{\mu \rho_\mathrm{i}}}{\sigma^2 - \omega_\mathrm{Ai}^2} \right)^2 \right].
\end{equation}

The equation governing plasma dynamics outside the jet (without twist and velocity field, \emph{i.e.}, $A = 0$, $\Omega = 0$, and $U_z = 0$) is the same Bessel equation, but $\kappa_\mathrm{i}$ is replaced by $k_z$.

Inside the jet, the solution to Equation~\ref{eq:bessel} bounded on the jet axis is the modified Bessel function of the first kind,
\begin{equation}
\label{eq:ptoti}
    p_\mathrm{tot}(r \leqslant a) = \alpha_\mathrm{i}I_m(\kappa_\mathrm{i}r),
\end{equation}
where $\alpha_\mathrm{i}$ is a constant.

Outside the jet, the solution bounded at infinity is the modified Bessel function of the second kind,
\begin{equation}
\label{eq:ptote}
    p_\mathrm{tot}(r > a) = \alpha_\mathrm{e}K_m(k_z r),
\end{equation}
where $\alpha_\mathrm{e}$ is a constant.

To obtain the dispersion equation governing the propagation of MHD modes along the jet, the solutions at the jet surface need to be merged through boundary conditions.

It is well known that for non-rotating and untwisted magnetic flux tubes the boundary conditions are the continuity of the Lagrangian radial displacement and total pressure perturbation at the tube surface \citep{chandrasekhar1961}, that is,
\begin{equation}
\label{eq:bc1}
    \xi_{\mathrm{i}r}|_{r=a} = \xi_{\mathrm{e}r}|_{r=a} \qquad \mbox{and} \qquad \left.p_\mathrm{tot\,i} \right\vert_{r=a} = p_\mathrm{tot\,e}|_{r=a},
\end{equation}
where total pressure perturbations $p_\mathrm{tot\,i}$ and $p_\mathrm{tot\,e}$ are given by Equations~\ref{eq:ptoti} and \ref{eq:ptote}, respectively.

When the magnetic flux tube is twisted and still non-rotating and the twist has a discontinuity at the tube surface, then the Lagrangian total pressure perturbation is continuous and the boundary conditions are (see, \emph{e.g.}, \citealp{bennett1999}; \citealp{zaqarashvili2010}; \citealp{zaqarashvili2014})
\begin{equation}
\label{eq:bc2}
    \xi_{\mathrm{i}r}|_{r=a} = \xi_{\mathrm{e}r}|_{r=a} \qquad \mbox{and} \qquad \left.p_\mathrm{tot\,i} - \frac{B_{\mathrm{ i}\phi}^2}{\mu a}\xi_{\mathrm{i}r}\right\vert_{r=a} = p_\mathrm{tot\,e}|_{r=a}.
\end{equation}
The second term in the second boundary condition stands for the pressure from the magnetic tension force.

If a non-twisted ($B_{\mathrm{i}\phi} = 0$) tube rotates and the rotation has discontinuity at the tube surface, the boundary condition for the Lagrangian total pressure perturbation has a similar form as the second boundary condition in Equation~\ref{eq:bc2}
\begin{equation}
\label{eq:bc3}
    \xi_{\mathrm{i}r}|_{r=a} = \xi_{\mathrm{e}r}|_{r=a} \qquad \mbox{and} \qquad \left.p_\mathrm{tot\,i} + \frac{\rho_\mathrm{i} U_{\phi}^2}{a}\xi_{\mathrm{i}r}\right\vert_{r=a} = p_\mathrm{tot\,e}|_{r=a}.
\end{equation}
Here, the second term that describes the contribution of centrifugal force to the pressure balance can be derived from Equation~\ref{eq:xi_r-new} for $B_{\mathrm{i}\phi} = 0$ by multiplying that equation by $\mathrm{d}r$, and considering the limit of $\mathrm{d}r \to 0$ through the boundary $r = a$, one obtains the relation $\mathrm{d}\left[ p_\mathrm{tot} + (\rho_\mathrm{i} U_\phi^2/a)\xi_r \right] = 0$, or, equivalently, the second boundary condition in the above equation.  Hence, the boundary condition for the Lagrangian total pressure perturbation in rotating and magnetically twisted flux tubes has the form
\begin{equation}
\label{eq:bc4}
    \left.p_\mathrm{tot\,i} + \left( \frac{\rho_\mathrm{i} U_{\phi}^2}{a} - \frac{B_{\mathrm{i}\phi}^2}{\mu a} \right)\xi_{\mathrm{i}r}\right\vert_{r=a} = p_\mathrm{tot\,e}|_{r=a}.
\end{equation}
In the case of uniform rotation and magnetic field twist, Equation~\ref{eq:uniform}, the boundary conditions for Lagrangian radial displacement $\xi_r$ and total pressure perturbation $p_\mathrm{tot}$ are
\begin{equation}
\label{eq:bc5}
    \xi_{\mathrm{i}r}|_{r=a} = \xi_{\mathrm{e}r}|_{r=a} \quad \mbox{and} \quad \left.p_\mathrm{tot\,i} + a\left( \rho_\mathrm{i} \Omega^2 - \frac{A^2}{\mu} \right)\xi_{\mathrm{i}r}\right\vert_{r=a} = p_\mathrm{tot\,e}|_{r=a}.
\end{equation}
Using these boundary conditions we obtain the dispersion equation of normal MHD modes propagating in rotating and axially moving twisted magnetic flux tubes \citep{zaqarashvili2015}
\begin{eqnarray}
\label{eq:dispeq-new}
    \frac{\left( \sigma^2 - \omega_\mathrm{Ai}^2 \right)F_m(\kappa_\mathrm{i}a) - 2m\left( \sigma \Omega + A\omega_\mathrm{Ai}/\! \sqrt{\mu \rho_\mathrm{i}} \right)}{\rho_\mathrm{i}\left( \sigma^2 - \omega_\mathrm{Ai}^2 \right)^2 - 4\rho_\mathrm{i}\left( \sigma \Omega + A\omega_\mathrm{Ai}/\! \sqrt{\mu \rho_\mathrm{i}} \right)^2} \nonumber \\
    \nonumber \\
    {}= \frac{P_m(k_z a)}{\rho_\mathrm{e}\left( \sigma^2 - \omega_\mathrm{Ae}^2 \right) - \left( \rho_\mathrm{i}\Omega^2 - A^2/\mu \right)P_m(k_z a)},
\end{eqnarray}
where
\[
    F_m(\kappa_\mathrm{i}a) = \frac{\kappa_\mathrm{i}aI_m^{\prime}(\kappa_\mathrm{i}a)}{I_m(\kappa_\mathrm{i}a)}, \quad P_m(k_z a) = \frac{k_z aK_m^{\prime}(k_z a)}{K_m(k_z a)}, \quad \omega_\mathrm{Ae} = \frac{k_z B_\mathrm{e}}{\sqrt{\mu \rho_\mathrm{e}}}.
\]

We note that in the case of non-rotating twisted flux tube ($\Omega = 0$) above Equation~\ref{eq:dispeq-new} recovers the well-known dispersion relation of normal MHD modes propagating in cylindrical twisted jets (see, \emph{e.g.}, \citealp{zhelyazkov2012a}).  If the environment medium is a cool plasma as is the case of our macrospicule, where the thermal pressure $p_\mathrm{e} = 0$, the $k_z a$ in $P_m(k_z a)$ must be replaced by $k_z a[ 1 - \left( \omega/\omega_\mathrm{Ae} \right)^2 ]^{1/2}$, which yields the wave dispersion relation we used in Equation~\ref{eq:dispeq}.

%
\begin{acks}
Our work was supported by the Bulgarian Science Fund under project DNTS/INDIA 01/7.  The authors are indebted to the anonymous reviewer for pointing out a mathematical error and for the helpful and constructive comments and suggestions that contributed to improving the final version of the manuscript.

\medskip
\noindent
\textbf{Disclosure of Potential Conflicts of Interest:} The authors declare that they have no conflicts of interest.
\end{acks}

%
%
%

\end{article}
\end{document}